\documentclass[12pt]{article}
\usepackage{latexsym,epsfig,citesort,graphics}

\newcommand{\be}{\begin{equation}}
\newcommand{\ee}{\end{equation}}
\newcommand{\bq}{\begin{eqnarray}}
\newcommand{\eq}{\end{eqnarray}}
\newcommand{\beq}{\begin{equation}}
\newcommand{\eeq}{\end{equation}}
\newcommand{\bea}{\begin{eqnarray}}
\newcommand{\eea}{\end{eqnarray}}
\begin{document}
\begin{titlepage}
\vskip-.5in
\begin{flushright}
{\small LBNL-51871 \\
UCB-PTH-02/58 \\
UFIFT-HEP-02-36}
\end{flushright}

\vskip .5in
\begin{center}
{\large \bf An Improved Mean Field Approximation on the\\ 
Worldsheet for Planar $\phi^3$ Theory}
\footnote{This work was supported in part
 by the Director, Office of Science,
 Office of High Energy and Nuclear Physics, 
 of the U.S. Department of Energy under Contract 
DE-AC03-76SF00098, in part by the National Science Foundation Grant
22386-13067-44-X-PHHXM, and in part by the Department of Energy
under Grant No. DE-FG01-97ER-41029}
\vskip .50in


Korkut Bardakci$^a$ and Charles B. Thorn$^b$
\vskip.1in
{\small\em $^a$Department of Physics,
University of California at Berkeley\\
\vskip.05in
$^a$Theoretical Physics Group,
    Lawrence Berkeley National Laboratory\\
      University of California,
    Berkeley CA 94720\\
\vskip.05in
$^b$Institute for Fundamental Theory,
Department of Physics\\ 
University of Florida, Gainesville FL 32611}
\end{center}

\vskip .3in

\begin{abstract}
We present an improved version of our earlier work on summing the
planar graphs in $\phi^{3}$ field theory. The present treatment
is also based on our world sheet formalism and the mean field 
approximation, but it makes use of no further approximations.
We derive a set of equations between the expectation values of
the world sheet fields, and we investigate them in certain
limits. We show that the equations can give rise to 
(metastable) string forming solutions.
\end{abstract}
\end{titlepage}

\newpage
\renewcommand{\thepage}{\arabic{page}}
\setcounter{page}{1}

\noindent{\bf 1. Introduction}
\vskip 9pt

One of the most challenging problems in the study of non-Abelian
gauge theories is to discover the dual string description 
\cite{gubserkpanomalous}.
 After many years of relative stagnation,
the discovery of the AdS/CFT correspondence \cite{maldacena,aharonygmoo}
has led to important progress in the resolution of this problem. 
This approach relies heavily on ideas from string theory and supersymmetry,
and it can deal effectively with only a fairly restricted class of field
theories.

Recently, we initiated a program \cite{bardakcit}, 
one of the goals of which is to establish the duality between
field and string theories. The idea, which goes back to
Nielsen, Olesen, Sakita and Virasoro
\cite{nielsenfishnet}, and which was given a systematic foundation via
't Hooft's $1/N_c$ expansion 
\cite{thooftlargen}, 
is to construct a world sheet description of the planar graphs of a
general field theory. As shown in \cite{thooftlargen}, 
planar graphs are selected
by introducing an internal color degree of freedom and taking
$N_{c}\rightarrow\infty$ limit, and the use of a 
mixed coordinate-momentum space light cone coordinates 
leads to the world sheet description.
 Although originally only a $\phi^{3}$ theory was considered for simplicity
\cite{bardakcit,thooftlargen}, 
this approach is sufficiently flexible to be able to
handle more realistic cases, including nonsupersymmetric 
\cite{thornsheet} and supersymmetric \cite{gudmundssontt} gauge theories.

The original world sheet model suggested by 't Hooft was non-local. In
reference \cite{bardakcit}, it was shown how to reformulate this model
as a local world sheet field theory by introducing additional
non-dynamical fields. The world sheet structure of a given
 graph consists of the bulk and a bunch of ``solid''
lines, which describe the multiple internal boundaries
representing the loops of a multi-loop planar diagram. 
The dynamics
is all in the boundaries, and the only function of the
 fields that live in the bulk is to instantaneously
transmit the interaction from one boundary to another.
Therefore, the local world sheet theory can be thought of as a
topological theory.

The representation of planar graphs by a topological field
theory is perhaps elegant, but it has no new physical content.
The elimination of the bulk fields would lead back to the
original non-local structure of 't Hooft. One gets something
new and interesting if the topological bulk fields are somehow promoted into
genuine dynamical fields. One way for this to happen is through
the condensation of the solid lines (the boundaries).
In a graph of a given order, the density of the solid lines, by which
we mean the percentage of the area on the world sheet occupied by
solid lines, is zero. This is because solid lines are lines; they have
zero thickness. However, as the order of the graph asymptotes to
infinity, solid lines become more and more numerous and dense, and
ultimately, one can envisage a limit in which they acquire a finite
density on the world sheet. This is what we mean by the condensation
of the boundaries. In this limit, the distinction between the bulk
and the boundary disappears,  the world sheet acquires a uniform
texture, and it becomes possible to have
 a string formulation of the sum of the Feynman graphs
in terms of dynamical fields on the world sheet.
We will call this mechanism string formation through
condensation of boundaries.
Whether this really happens depends on the dynamics: One clearly
needs an attractive interaction and also an effectively strong coupling,
leading to the domination of higher order graphs. 

The worldsheet formalism of \cite{bardakcit} might also 
provide a natural
setting for understanding confinement in real QCD, where 
strong coupling is only a feature of infrared dynamics. Then
the physical world sheet representing the confining flux tube
would not have the uniform texture mentioned above but would
include important fluctuations in which regions of the world
sheet would not have solid line condensation and would retain
the topological character of perturbation theory. These fluctuations would 
describe the point-like structures implied by asymptotic freedom.

In an earlier paper \cite{bardakcitmean},  using
$\phi^{3}$ theory as a prototype toy model, we investigated the possibility
of boundary condensation leading to string formation. Although this is an
unphysical theory, it provides a simple setting for developing
the tools needed to attack more interesting and also more
complicated theories. The main tool used in \cite{bardakcitmean}
was the mean field approximation, or the self consistent
field approximation. This approximation
 has been widely used in both field theory and many body
theory, in most cases leading to at least qualitatively reasonable
answers. For the problem at hand, the easiest way to implement
it systematically is to consider the limit of large $D$, where
$D$ is the number of transverse dimensions in the light cone
picture. We should add that although the large $D$ limit
provides a convenient bookkeeping device, it does not 
truly capture the physics of the problem in most cases.
The critical factor is the number of degrees of freedom
of the system in question: The mean field approximation
is usually successful when applied to a system with a large
number of degrees of freedom. Of course, one way to to have a large
number of degrees of freedom is to have a large number of
space dimensions, but this is rather artificial in most
cases. A more physical situation is to have a large number
of particles in a many body system, and this is the case
in most of the standard applications. In our case, as explained
above, we are interested in graphs with a large number of
boundaries, and we can roughly identify the number of 
boundaries with the degrees of freedom. With this identification,
the mean field method should be a good tool for investigating
problems involving graphs with a large number of boundaries.
There is then no need to appeal to the large $D$ limit, except,
as a convenient bookkeeping device.

In this article we aim to improve the treatment of
\cite{bardakcitmean} in several respects. First, we
applied the mean field approximation to a system of Ising
spins introduced as a representation of the sum over all
arrangements of solid lines. Since the world sheet on which these 
spins lived remained  2 dimensional even as $D\to\infty$,
it was not clear why that should justify replacing each
spin by a mean field. Instead the large $D$ limit is
actually that of an $O(D)$ vector model. As is well-known this
limit justifies treating the $O(D)$ invariants, the scalar
products of the vector fields, classically. In fact the process
of replacing the scalars with classical fields eventually
leads to a mean field treatment of the Ising spins as well,
but, unlike in \cite{bardakcitmean}, there are no ambiguities
in how to set up the action for the classical mean fields.
We incorporate this clarification in the current article.
But the most substantial improvement we make here is in the
treatment of the Dirichlet boundary conditions. In \cite{bardakcitmean}
we introduced Gaussian representations of delta functions of the
type
$$
\delta(\dot{\bf q})=\lim_{\beta\to\infty}\left({\beta\over2\pi}\right)^{D/2}
e^{-\beta{\dot{\bf q}^2}}\nonumber
$$
to enforce the boundary conditions. Keeping $\beta$
finite amounts to imposing an infrared cutoff on target space.
With a sufficiently exact
calculation there is nothing wrong with this. But 
after making an approximation to the cutoff theory, the
delicacy of removing the infrared cutoff might well invalidate
the approximation. In the current article we shall impose
the boundary conditions exactly, thereby eliminating this
difficulty. The price will be the need to introduce a more complicated
system of mean fields, some of which are nonlocal on the world sheet.
The presence of such nonlocal mean fields might even make
it easier to incorporate the non-homogeneous worldsheet
textures required for a realistic worldsheet description
of QCD.
We are then able to derive a set of equations for
these fields, which relies on
no other approximation except for the mean field approximation.
We consider these equations (eqs.(46) and (47)) as the main result
of this article. Although these equations are non-linear and 
somewhat complicated, it should be possible to attack them
by various approximation schemes and numerical methods.
However, in this article, we will be content with studying them
in some limiting cases,
and we leave a full investigation of this problem to future research.

The paper is organized as follows. In the next section, we present
 a brief review of the world sheet formalism developed in \cite{bardakcit} 
for summing planar graphs. We also argue that, in the leading order of
the mean field approximation, a good deal of simplification takes place:
The ghost fields can be dropped, and the world sheet dynamics is
described by a free action (eq.(7)), plus constraints on the
boundaries given by eq.(8). In section 3, the boundary constraints
are re-expressed in a form suitable for inclusion in the effective
action and also for taking the large $D$ limit.
This is done by introducing auxiliary scalar fields, which are
equal to the scalar products of the vector fields in $D$ dimensions.
The important point is that these scalar fields become classical
in the large $D$ limit. Section 4 is devoted to the construction
of the effective action as a function of the  scalar
fields. This is done by explicitly carrying out the
functional integration over the vector fields. The classical equations
for the scalar fields resulting from this effective action
completely determines the large $D$ dynamics. In section 5, we 
investigate these equations in various limits, and we argue that
one solution, although not stable,
can lead to condensation of the
boundaries and string formation.
 This is  a moderately encouraging but
preliminary result, although no definite conclusions can be
reached until we have the full solution.
 Section 6 summarizes our conclusions.
In the appendix, we show that the contribution of the ghost
fields is unimportant in the leading order of the large $D$
limit.

\vskip 9pt
\noindent {\bf 2. A Brief Review }
\vskip 9pt

We will be working with a massless $\phi^{3}$ matrix field theory
in the large $N$ limit, which amounts to a summation of  the
planar graphs. As 't Hooft has shown \cite{thooftlargen}, Feynman rules are
especially simple if a particular mixture of coordinate and
momentum light cone variables are used. We will use the following
notation:
 A Minkowski vector $v^{\mu}$
will be written as $(v^{+}, v^{-},{\bf v})$, where
$v^{\pm}=(v^{0} \pm v^{3})/\sqrt{2}$, and the boldface letters label the
components along the transverse directions. The Lorentz invariant
product of two vectors $v$ and $w$ is given by $v\cdot w=
{\bf v}\cdot {\bf w}- v^{+} w^{-} -v^{-} w^{+}$. 
The evolution parameter (time) is $x^{+}$,
 and the Hamiltonian conjugate to
this time is $p^{-}$. A massless on-shell particle thus has the
energy $p^{-}= {\bf p}^{2}/2 p^{+}$.
 The number
of transverse dimensions $D$ is arbitrary to start with, and eventually,
we will consider the large $D$ limit. As explained in \cite{bardakcit}, this
limit is a convenient method of organizing the mean field
approximation. Of course, the eventual case of interest is $D=2$.

Let us now briefly review the Feynman rules derived in \cite{thooftlargen}. 
A specific graph is represented by a set of parallel solid lines
drawn on the world sheet.  A propagator corresponds
to a strip bounded by two solid lines (Fig.(1)).
\begin{figure}[t]
\centerline{\epsfig{file=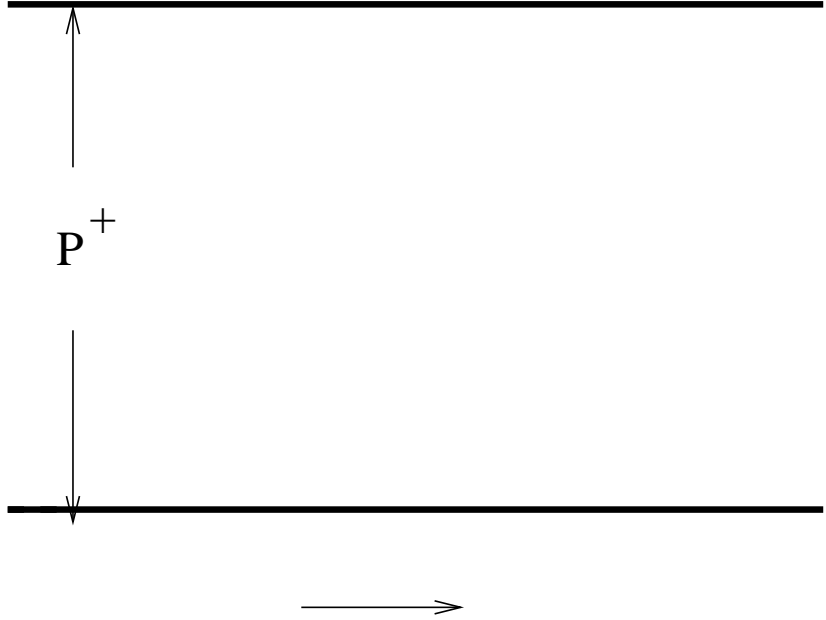,width=12cm}}
\caption{A Propagator}
\end{figure}
 If $p$ is the
momentum carried by the propagator, the strip has width $p^{+}$,
and length $\tau= x^{+}$.
We associate two transverse
momenta ${\bf q}_{1}$ and ${\bf q}_{2}$ with the two solid lines 
forming the boundary.  The transverse
momentum ${\bf p}$ of the propagator is their difference:
$$
{\bf p}= {\bf q}_{1}- {\bf q}_{2},
$$
and using a Euclidean world sheet metric, the propagator is given by
\be
\frac{\theta(\tau)}{2 p^{+}} \exp\left(-  \tau\, ({\bf q}_{1}
-{\bf q}_{2})^{2}/ 2 p^{+}\right),
\ee

Now let us consider a more complicated
graph, with interaction vertices,
 pictured in Fig.(2).
\begin{figure}[t]
\centerline{\epsfig{file=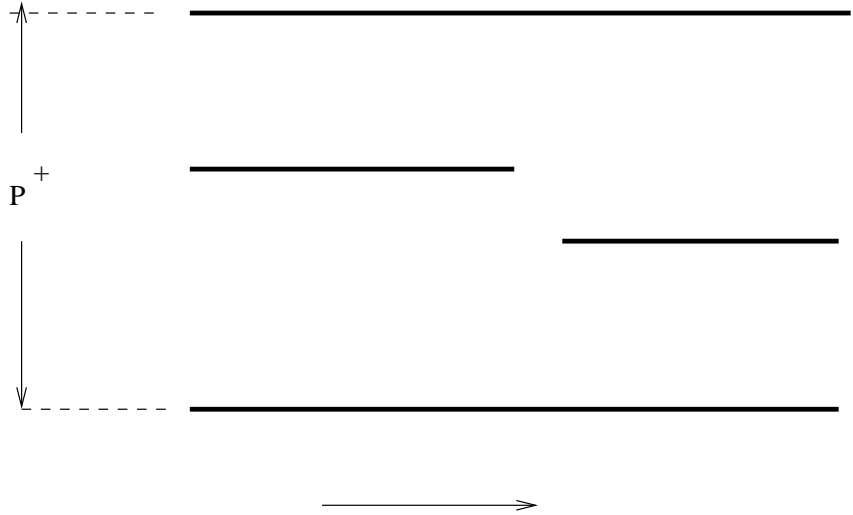,width=12cm}}
\caption{A Typical Graph}
\end{figure}
 Interaction takes place at  the points
where a solid line ends, and a factor of $ g$ is associated with
each such vertex, where $g$ is the coupling constant. After
putting together the propagators and coupling constants associated
with the graph, one
has to integrate over the positions of the solid lines and the position
of the interaction vertices, as well as the momenta carried by solid
lines. We note that momentum conservation
is automatic in this formalism.

These Feynman rules look quite non-local; however,
it was shown in reference \cite{bardakcit} that they can be
reproduced by a local field theory defined on the world sheet.
Later, this result was extended from $\phi^{3}$
 to more complicated and more
realistic  theories \cite{thornsheet,gudmundssontt}. The world sheet 
 field theory in question can be
formulated either on a latticized or continuum world sheet,
and here we choose the continuum version. We will eventually
need to specify some cutoffs, but we defer that until the
need arises. Using the coordinates
 $\sigma$ in the $p^{+}$ direction
and  $\tau$ itself in the $x^{+}=\tau$ direction,  the
bosonic fields ${\bf q}(\sigma,\tau)$ and the fermionic fields
$b(\sigma,\tau)$ and $c(\sigma,\tau)$, called ghosts,
are introduced. In contrast
to ${\bf q}$, which has $D$ components, $b$ and $c$ each have $D/2$ 
components. (Here, we are assuming that $D$ is even). The free part of
the action on the world sheet is given by
\be
S_{0}= \int_{\tau_{i}}^{\tau_{f}} d\tau \int_{0}^{p^{+}}
d \sigma \left(b'\cdot c' - \frac{1}{2} {\bf q}'^{2}\right),
\ee
where the prime denotes derivative with respect to $\sigma$.
Here $p^{+}$ is the $+$ component of the total momentum flowing
through the graph. For convenience, we have also restricted
the $\tau$ integration; eventually, we will let $\tau_{i}
\rightarrow -\infty$ and $\tau_{f}\rightarrow +\infty$.
This action is to be supplemented by the following boundary conditions:
On  solid lines, ${\bf q}(\sigma,\tau)$ is constrained to be
independent of $\tau$, or, equivalently, the Dirichlet boundary
condition
$$
\dot{{\bf q}}=0
$$
is imposed, where the dot denotes derivative with respect to $\tau$.
Also, if the transverse momentum carried by the whole graph is
${\bf p}$, the constraint
\be
\int _{0}^{p^{+}} d\sigma\, {\bf q}' = {\bf p}
\ee
has to be imposed. For simplicity, we will set ${\bf p}=0$ in what
follows, which means periodic boundary conditions on ${\bf q}$:
\be
{\bf q}(\sigma=0)={\bf q}(\sigma=p^{+}).
\ee
The corresponding boundary conditions on the ghosts are simple:
\be
b=c=0
\ee
on the solid lines.

In addition to $S_{0}$, it was shown in \cite{bardakcit} that
 the full action contains interaction
terms in the form of the insertion of ghost vertices. These insertions
are needed to correctly reproduce the factor $1/2 p^{+}$
in front of the exponential in the expression (1) for the
propagator. If, for example, one integrates the 
matter and ghost fields
of $S_{0}$ with the boundary conditions appropriate for the
propagator, without any vertex ghost insertions, the result is
\be
\theta(\tau)\exp\left(-\tau\, {\bf p}^{2}/2 p^{+}\right),
\ee
instead of (1). Local ghost vertex insertions can be arranged
to correct this defect. However,
 we will now argue that this is the correct leading term in the
large $D$ expansion. To see
this, let us rewrite the propagator in the form
$$
\theta(\tau)\exp\left(-\tau\,{\bf p}^{2}/2 p^{+} -\ln( 2 p^{+})
\right).
$$
The first term in the exponent really corresponds to $D$ terms
since it is the square of a vector with $D$ components.
Consequently, in the large $D$ limit, its contribution is of
the order of $D$. In contrast, the contribution of the second
term is $D$ independent. Therefore, in the leading order in $D$,
which is all we are going to consider in this paper, 
 the factor of $1/2 p^{+}$ can be dropped. This means that, in
the large $D$ limit, we are entitled to work with $S_{0}$ given
by eq.(2) and ignore the ghost insertion terms.

We are going to take advantage of one more simplification.
Although the fermionic (ghost) part of $S_{0}$ does contribute
in the large $D$ limit, it will be shown in the appendix that its
contribution harmlessly shifts some of the fields
we had already introduced. So, bearing in mind
that the fields might need to be interpreted differently, we are going to
drop the ghost contributions.

 To summarize, the starting point of the present work
will be the action
\be
S_{0}\rightarrow - \int_{\tau_{i}}^{\tau_{f}} d\tau
\int_{0}^{p^{+}} d \sigma\left(\frac{1}{2} {\bf q}'^{2}
\right),
\ee
plus  the Dirichlet boundary conditions
\be
\dot{{\bf q}}=0
\ee
on solid lines and the periodicity condition given by eq.(4).
We would like to stress that the terms we have neglected
do contribute in the non-leading orders of the large $D$ expansion.

\vskip 9pt
\noindent{\bf 3. The Boundary Conditions}
\vskip 9pt

Our goal is to express the sum over all the planar graphs in the
form of an effective action. We start with eqs.(7) and (8),
which generate the simplified Feynman rules of the large $D$ limit.
We would like to
rewrite the  conditions (8) in a form that can be identified
as a contribution to the action, and cast them
in a form suitable for taking the large $D$ limit.
 We note that the
boundary conditions decouple in the $\sigma$ direction, and so it
is convenient first to discretize the $\sigma$ coordinate
 into small segments of length $\Delta\sigma=\epsilon_{1}$,
and to impose the conditions for each discrete value of 
$\sigma$ separately. (In \cite{bardakcitmean} this cutoff $\epsilon_1$
was called $m$ the minimal unit of $p^+$ in the
discretized formulation.) The parameter $\epsilon_{1}$ will play the role
of an ultraviolet cutoff on the worldsheet. 
It should be pointed out that we are
making an important change in the way we classify graphs. The
customary classification is according to the total number of
solid lines, which is the same as counting the powers of the
coupling constant. This is analogous to a Fock space description.
Instead, we are now going to focus on the distribution of solid lines
separately for each value of $\sigma$. This is more analogous
to the occupation number description. If one is interested in the
condensation of the solid lines, the occupation number description
is clearly superior, since counting the number of lines in
a condensate is not very useful.

Next, for a fixed value of $\sigma$, we are going to  carry out 
the summation over all possible partitionings of solid lines and
over all values of momenta flowing through them. This will result
in an integral equation.
 In particular, we note that the Dirichlet boundary conditions
(8) will be taken into account exactly. In this respect, the present
treatment is superior to the one given in \cite{bardakcitmean}; in that work,
only a weaker version of these boundary conditions was imposed.

In Fig.(3), we have drawn lines located at the discrete values
of $\sigma$ and extending in the $\tau$ direction.
\begin{figure}[t]
\centerline{\epsfig{file=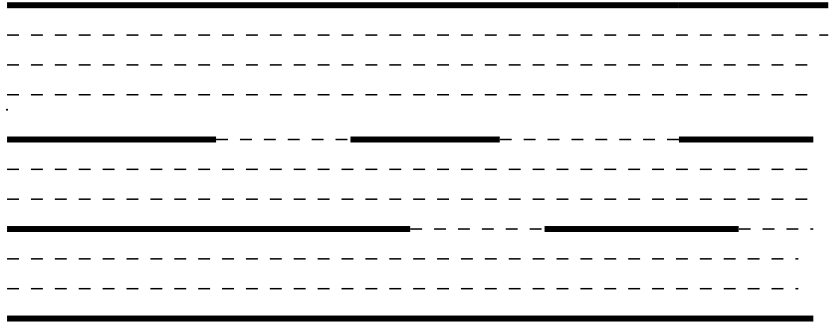,width=12cm}}
\caption{Solid and Dotted Lines}
\end{figure}
 These lines consist
of alternating solid and dotted segments:
 As before, the solid lines are located where
the boundary conditions (8) hold, whereas the dotted
 lines run through what
used to be blank space, where there are no boundary conditions.
 Now consider
a typical line located at some $\sigma$, starting at $\tau_{i}$ and
ending at a variable point $\tau$. For simplicity, we assume that
this line contains an equal number of solid and dotted line segments,
and we call this number $m$.
 One has then to sum
over all possible partitions of this line between alternating
solid and dotted segments for a given $m$, and then sum over all $m$.
 Let $F_{m}(\sigma,\tau_{i},\tau)$ be the factor that takes care
of  (8) on the solid segments.  This
function satisfies the following recursion relation:
\be
F_{m+1}(\sigma,\tau_{i},\tau)= g^{2}\int_{\tau_{i}}^{\tau}
d\tau_{1} \int_{\tau_{1}}^{\tau} d\tau_{2}\, K(\sigma,\tau_{2},\tau)
F_{m}(\sigma,\tau_{i},\tau_{1}).
\ee
In going from $F_{m}$ to $F_{m+1}$, we have added a dotted
segment extending from $\tau_{1}$ to $\tau_{2}$, and a solid
segment extending from $\tau_{2}$ to $\tau$. Since there is
no boundary condition on the dotted segment,
 we associate with it a factor
of one. With the solid segment, we associate the factor 
$K(\sigma,\tau_{2},\tau)$, which is introduced to take care
of the corresponding boundary condition.
 The total $F$ is then given by
\be
F(\sigma,\tau_{i},\tau)= \sum_{m=0}^{\infty}
 F_{m}(\sigma,\tau_{i},\tau).
\ee
and it satisfies the integral equation
\be
F(\sigma,\tau_{i},\tau)= F_{0}(\sigma,\tau_{i},\tau)
+ g^{2} \int_{\tau_{i}}^{\tau} d\tau_{1} \int_{\tau_{1}}
^{\tau} d\tau_{2}\, K(\sigma,\tau_{2},\tau) F(\sigma,\tau_{i},
\tau_{1}).
\ee

So far, we have defined an $F$ that implements (8)
for a single line located at $\sigma$. To find the $F$ that
implements these boundary conditions
 for the whole world sheet, we have 
to multiply the $F$'s associated with each discrete value of
$\sigma$ and let $\tau\rightarrow \tau_{f}$:
\be
F(\tau_{i},\tau_{f})=\prod_{n} F(\sigma_{n},\tau_{i},\tau_{f}).
\ee

Now let us return to the integral equation (11). 
We will set
\be
F_{0}(\sigma,\tau_{i},\tau)= \delta(\tau_{i}-\tau).
\ee
The kernel $K$, which has to implement (8) on the solid
line segment extending from $\tau_{2}$ to $\tau$, is given by
\be
K(\sigma,\tau_{2},\tau)=\int d{\bf k} \int D{\bf l}\,
 \exp\left(i\int_{\tau_{2}}^{\tau} d\tau'\, {\bf l}(\sigma,\tau')
\cdot \Big({\bf q}(\sigma,\tau') - {\bf k}\Big)\right).
\ee
The integration over ${\bf l}(\sigma,\tau')$ sets
 ${\bf q}(\sigma,\tau')$
equal to a constant $\tau$ independent vector ${\bf k}$ in the
interval $(\tau_{2},\tau)$, and then ${\bf k}$ is integrated over.
It is clear that this is equivalent to the condition that
${\bf q}$ is independent of $\tau'$ in this interval.

It will become clear as we proceed that an ultraviolet
cutoff in the variable ${\bf k}$ is needed to avoid
divergences. We therefore modify the expression for $K$
by introducing a suitable cutoff of the form of
$\exp(-\epsilon k^{2})$ and then do the integration over
${\bf k}$:
\bq
&&K(\sigma,\tau_{1},\tau_{2})\rightarrow\int d{\bf k}
\int D{\bf l}\,\exp\left(-\epsilon k^{2} + i
\int_{\tau_{1}}^{\tau_{2}} d\tau' {\bf l}(\sigma,\tau')
 \cdot \Big({\bf q}(\sigma,\tau') - {\bf k}\Big)\right)
\nonumber\\
&&=\int D{\bf l}\,
 (\pi/\epsilon)^{D/2} \exp\left(i\int_{\tau_{1}}^{\tau_{2}}
d\tau' {\bf l}(\sigma,\tau')\cdot {\bf q}(\sigma,\tau')\right.\nonumber\\
&&\left.\qquad
 -\frac{1}{4\epsilon}
\int_{\tau_{1}}^{\tau_{2}} d\tau' \int_{\tau_{1}}^{\tau_{2}}
d \tau'' {\bf l}(\sigma,\tau')\cdot {\bf l}(\sigma,\tau'')
\right).
\eq
We note that in addition to the worldsheet
 ultraviolet cutoff $\epsilon_{1}$
introduced earlier, we now have a second ultraviolet cutoff
$\epsilon$, acting in the target space.
Later, we will find that a third cutoff may be necessary,
although it is likely that the third cutoff is
expressible in terms of the first two.

It is worth noting that the nonlocal term in eq.(15) is entirely
due to the integration over ${\bf k}$, the Dirichlet value
of ${\bf q}$ on the internal boundary which $K$ adds to the worldsheet.
But in the worldsheet formalism applied to gauge theories with extended 
supersymmetry \cite{gudmundssontt}, the worldsheet variables
${\bf q}(\sigma,\tau)$ were $2+d$ dimensional where $d=2$
for ${\cal N}=2$ supersymmetry and $d=6$ for ${\cal N}=4$ supersymmetry.
The process of dimensional reduction, which leads to the
4 dimensional theory with extended supersymmetry, was achieved 
in \cite{gudmundssontt} by
imposing the boundary condition $q^I=0$ on all boundaries, external
and internal, whenever $I=3,4,\ldots d$. This means that the
same components of the vector ${\bf k}$ are put to zero and {\it not}
integrated. In other words the nonlocality is not 
introduced for the components of ${\bf l}$ in the extra dimensions.
This worldsheet locality for the dimensions associated with the 
field theoretic internal degrees of freedom could be associated
with the generic appearance of $S_n$ factors of the space-time
manifold on the strong coupling side of the AdS/CFT 
correspondence\footnote{We thank J. Maldacena for an illuminating
conversation that inspired this suggestion.}.   

We are dealing here with a vector model, and the standard
first step in taking the large $N$ (in this case, the large
$D$) limit, is first to express everything in terms of
the scalar products of the vectors. From eq.(15), we see that
this goal has been achieved: The kernel
 $K$ is expressed
in terms of the scalar products ${\bf q}\cdot {\bf l}$ and
${\bf l}\cdot {\bf l}$. In the next step, the scalar products
 are replaced by their
vacuum expectation values, which are proportional to D, and
 which we denote by $D\, \phi_{1}$ and
$D\, \phi_{2}$ respectively:
\bq
{\bf l}(\sigma,\tau')\cdot {\bf q}(\sigma,\tau')&\rightarrow&
\langle {\bf l}(\sigma,\tau')\cdot {\bf q}(\sigma,\tau')
\rangle = D\, \phi_{1}(\sigma,\tau'),\nonumber\\
{\bf l}(\sigma,\tau') \cdot {\bf l}(\sigma,\tau'')
&\rightarrow& \langle {\bf l}(\sigma,\tau') \cdot {\bf l}
(\sigma,\tau'')\rangle = D\, \phi_{2}(\sigma,\tau',\tau'').
\eq
We would like to make it clear that $\phi_{1}$ and $\phi_{2}$
are fixed classical background
fields, not to be
integrated over. They are the saddle points that dominate the
large $D$ limit,
  to be determined by
minimizing the effective action. The general rule is that
 vector valued fields are quantum mechanical, and 
they are to be integrated over. In contrast, scalar valued
fields are classical background fields.

The kernel $K$ can now be written solely in terms of $\phi_{1}$
and $\phi_{2}$:
\bea
K(\sigma,\tau_{1},\tau_{2})&=&\nonumber\\
&&\hskip-.5in (\pi/\epsilon)^{D/2} \exp\left(
i D \int_{\tau_{1}}^{\tau_{2}} d \tau' \phi_{1}(\sigma,\tau')
-\frac{D}{4\epsilon}
\int_{\tau_{1}}^{\tau_{2}} d\tau' \int_{\tau_{1}}^{\tau_{2}}
d \tau'' \phi_{2}(\sigma, \tau', \tau'')\right).\nonumber\\
&&
\eea
Substituting this expression back into eq.(11), we have to solve
an integral equation to find $F(\tau_{i},\tau_{f})$. We will
reconsider this problem in the next section, after the
simplifications following from translation invariance are
 taken into account. Here we note that once $F$ is known,
the total action $S$, which incorporates the boundary condition (8) on
solid lines, is given by
\be
\exp( S)= \exp( S_{1}) F(\tau_{i},\tau_{f}),
\ee
where,
\bq
S_{1}&=&\int_{0}^{p^{+}} d\sigma
 \int d\tau \left(- \frac{1}{2} ({\bf q}'(\sigma,\tau))^{2}
+i \lambda_{1}(\sigma,\tau)\Big({\bf l}(\sigma,\tau)\cdot {\bf q}(\sigma,\tau)
- D\, \phi_{1}(\sigma,\tau)\Big)\right)\nonumber\\
&-&i \int_{0}^{p^{+}} d \sigma \int d\tau \int d \tau'
\lambda_{2}(\sigma,\tau,\tau')\Big({\bf l}(\sigma,\tau)\cdot
{\bf l}(\sigma,\tau') - D\, \phi_{2}(\sigma,\tau,\tau')\Big).
\eq
The first term on the right is $S_{0}$ (eq.(7)). In the rest of the
terms in $S_{1}$,
the Lagrange multipliers $\lambda_{1}$ and $\lambda_{2}$ enforce
the definition of $\phi_{1,2}$ given by (16).

\vskip 9pt
\noindent{\bf 4. The Effective Action}
\vskip 9pt

In this section, we will address the problem of
 the computation of the effective action.
 This will involve
\begin{description}
\item a) solving the integral equation (11) for $F$ and
\item b) doing the functional integrations over
 the vector-valued fields ${\bf q}$ and ${\bf l}$, keeping
all the scalar valued fields fixed.
\end{description}
We will be able to make only partial progress in solving eq.(11).
In contrast, the functional integrations over ${\bf q}$ and ${\bf l}$
can be carried out in closed form.
 Once the effective action is derived,
the classical fields $\phi_{1,2}$ and $\lambda_{1,2}$ are determined
by the equations
\be
\frac{\delta S_{e}}{\delta \phi_{1,2}}=0,\;\;\;
\frac{\delta S_{e}}{\delta \lambda_{1,2}}=0.
\ee

The calculations outlined above are greatly simplified by making 
use of translation invariances in both the $\sigma$ and $\tau$
directions on the world sheet. This symmetry means that various
background fields are either constants or functions of a single
variable:
\bq
\lambda_{1}(\sigma,\tau)\rightarrow \lambda_{1}, & &
\phi_{1}(\sigma,\tau)\rightarrow \phi_{1},\nonumber\\
\lambda_{2}(\sigma,\tau,\tau')\rightarrow \lambda_{2}(\tau -\tau'),
& & \phi_{2}(\sigma,\tau,\tau')\rightarrow \phi_{2}(\tau -\tau'),
\eq
where $\lambda_{1}$ and $\phi_{1}$ are independent of $\sigma$ and
$\tau$, and $\lambda_{2}$ and $\phi_{2}$ depend only on the
difference $\tau -\tau'$. Substituting this in eq.(20), we see that 
$S_{1}$ will be non-local in the $\tau$ coordinate. This  makes the problem
of finding the minimum of the effective action more difficult;
instead of an algebraic equation, we have to solve functional equations
in one variable.

Let us now go back to the integral equation (11), taking advantage
of the simplification resulting from translation invariance. First,
we note that the functions $K$ and $F$ simplify:
\be
K(\sigma,\tau_{1},\tau_{2})\rightarrow K(\tau_{2}-\tau_{1}),
\;\;\; F(\sigma,\tau_{1},\tau_{2})\rightarrow F(\tau_{2}-\tau_{1}).
\ee
Now, eq.(11) can be formally solved using Fourier transforms. We define
\be
\tilde{K}(\omega)=\frac{1}{2 \pi}\int_{0}^{\infty} d \tau\,
e^{i \omega \tau} K(\tau),\;\;\;\tilde{F}(\omega)=\frac{1}{2\pi}
\int_{0}^{\infty} d \tau\, e^{i \omega \tau} F(\tau).
\ee
The lower limits of integration start at zero, since both $F$ and $K$
are defined to be zero for negative values of the argument. The
solution to eq.(11) is then given by
\be
F(\tau)=\int d \omega\,e^{-i \omega \tau} \tilde{F}(\omega)=
\frac{1}{2\pi}\int d \omega\,e^{-i \omega \tau} \frac{\omega}
{\omega- 2 \pi i g^{2} \tilde{K}(\omega)},
\ee
where $\tilde{K}$ can be written in the form
\be
\tilde{K}(\omega)=\frac{1}{2\pi} (\frac{\pi}{\epsilon})^{D/2}
\int_{0}^{\infty} d\tau \exp\left(i \omega \tau+i D \phi_{1}\tau
- \frac{D}{4\epsilon} L(\tau)\right)
\ee
and where we have defined
\be
L(\tau_{2}-\tau_{1})=
\int_{\tau_{1}}^{\tau_{2}} d\tau' \int_{\tau_{1}}
^{\tau_{2}} d\tau'' \phi_{2}(\tau'' -\tau').
\ee
The contribution of $F$ to the effective action follows from
eq.(18):
\be
S= S_{1}+ S_{F}, \;\;\; S_{F}=  \ln\left(F(\tau_{f}-\tau_{i})
\right).
\ee

Now, let us examine $S_{F}$ in more detail. The time interval
$\tau_{f}-\tau_{i}$, which corresponds to $\tau$ in eq.(23), will
eventually tend to infinity. It is a standard result in Fourier
transform that this limit will be dominated by the lowest lying
singularity in the variable $\omega$ of the integrand in eq.(23), which
could be the tip of a cut or a pole. Since, being conjugate to
time, $\omega$ is energy, a cut corresponds to the continuum and the
pole to a bound state. We are particularly interested in bound
states, since they can lower the energy of the system and lead to
a non-trivial solution to our variational problem. Therefore, we
will assume that the lowest lying singularity is a pole\footnote{
In reference \cite{bardakcitmean}, 
the energy spectrum consisted of only discrete
states due to the infrared cutoff introduced in the
Gaussian representation of the delta functions imposing
Dirichlet boundary conditions.}, and we
will be looking for the zeroes of the denominator in eq.(24):
\be
\omega- 2\pi i g^{2} \tilde{K}(\omega)=0.
\ee
If $\omega_{0}$ is a solution to this equation, the corresponding
energy is given by
\be
E_{0}=i \omega_{0},
\ee
and the corresponding contribution to the action is
\be
S_{F}= - \frac{i}{\epsilon_{1}} p^{+} (\tau_{f}-\tau_{i})
\omega_{0},
\ee
where $\epsilon_{1}$ is the cutoff resulting from the
discretization of the interval along the $\sigma$ direction.
(See eq.(12)). At this point, we should  note that 
bound states cannot arise in perturbation theory, so our
treatment from now on is definitely non-perturbative.
There is also the question of the reality of $E_{0}$.
 In a stable theory, the spectrum should be
real and bounded from below; however, since we are dealing
with $\phi^{3}$, an intrinsically unstable theory, we may
end up with complex energies. We will investigate this
question in section 5 in the context of a simple model.

For the time being, we cannot proceed any further with the
bound state problem without
knowing $\phi_{2}$, so we will leave its solution in the
implicit form defined by eqs.(24), (25) and (28).
Let us  now turn to the next problem, that
of carrying out the functional integrals over the quantum
fields ${\bf q}$ and ${\bf l}$.
  Replacing $\lambda_{1}$
and $\lambda_{2}$ in eq.(19)
 by their translation invariant forms (eq.(21)), this equation
can be rewritten as
\bq
S_{1}&=&S'_{1}+S''_{1},\nonumber\\
S'_{1}&=&i D \int_{0}^{p^{+}} d\sigma\,\left(-\int d\tau\,\lambda_{1}
\phi_{1}+\int d\tau_{1}\int d\tau_{2}\,\lambda_{2}(\tau_{1}-\tau_{2})
\phi_{2}(\tau_{1}-\tau_{2})\right),\nonumber\\
S_{1}''&=&\int_{0}^{p^{+}} d\sigma \int d\tau\left(-\frac{1}{2}
{\bf q}'^{2}+i \lambda_{1} {\bf l}\cdot {\bf q}\right)\nonumber\\
&-&i \int_{0}^{p^{+}} d\sigma \int d\tau_{1}\int d\tau_{2}\,
\lambda_{2}(\tau_{2} -\tau_{1})\,
{\bf l}(\sigma,\tau_{1})\cdot {\bf l}(\sigma,\tau_{2}).
\eq
It turns out to be more convenient to work in the momentum space.
We define
\bq
{\bf q}(\sigma,\tau)&=&\int dp_{0} \sum_{p_{1}}\exp(i p_{1}
\sigma+ i p_{0} \tau)\, \tilde{q}(p_{0},p_{1}),\nonumber\\
{\bf l}(\sigma,\tau)&=&\int d p_{0}\sum_{p_{1}}
\exp(i p_{1} \sigma +i p_{0} \tau)\, \tilde{l}(p_{0}, p_{1}),
 \nonumber\\
\lambda_{2}(\tau)&=&\int dp_{0}\,\exp(i p_{0}\tau)\,\tilde{\lambda}_
{2}(p_{0}),\nonumber\\
\phi_{2}(\tau)&=&\int d p_{0}\,\exp(- i p_{0} \tau)\,
\tilde{\phi}_{2}(p_{0}),
\eq
where
$$
p_{1}=2\pi m/p^{+},\;\; m\in Z,
$$
because of the periodic boundary conditions in the $\sigma$ direction.

Before attempting to do the functional integrals over ${\bf q}$ and 
${\bf l}$, we will examine the classical equations of motion. They
will enable us to make a comparison with the standard string action.
Written in momentum space, the equations of motion are
\bq
i \lambda_{1} \tilde{l} - p_{1}^{2} \tilde{q}&=&0,\nonumber\\
\lambda_{1} \tilde{q} - 4\pi \tilde{\lambda}_{2}(p_{0})
\tilde{l}&=&0.
\eq
The solution to these equations is
\be
p_{1}^{2}=\frac{i \lambda_{1}^{2}}{4\pi \tilde{\lambda}_{2}(
p_{0})}.
\ee
The usual string action corresponds to taking
\be
\frac{\lambda_{1}^{2}}{\tilde{\lambda_{2}}(p_{0})}
=16\pi i \alpha^{2}p_{0}^{2},
\ee
where $\alpha$ is the slope parameter. One can see this as
follows: In coordinate space, eq.(35) is equivalent to the action
$$
S_{s}= -\frac{1}{2}\int_{0}^{p^{+}} d\sigma \int d \tau
\left(4\alpha^{2}\dot{{\bf q}}^{2} + {\bf q}'^{2}\right).
$$
The Hamiltonian corresponding to this action is the light cone
energy $p^{-}$, and the quantized values of $p^{-}$ are given
by the usual string result
$$
p^{-}= \frac{ n\pi}{\alpha p^{+}},
$$
where $n$ is a positive integer. We are really interested in the
squares of the masses of the excitations, and these are given by
$$
M^{2}= 2 p^{+} p^{-}=  \frac{2n\pi}{\alpha},
$$
since we have set (see eq.(4)) ${\bf p}=0$.
From this result, we see that the slope parameter is given by 
$\alpha=(2T_0)^{-1}=\pi\alpha^\prime$, where $\alpha^\prime$
is the slope of open string Regge trajectories.
A string that is linearly 
confining at large distances corresponds to a $\tilde{\lambda}_{2}$
that has the dependence given by the  eq.(35) for small
$p_{0}$, so we are going to investigate whether such a behavior
is compatible with our dynamical scheme.

After this diversion,
 let us now carry out the integrations over $\tilde{q}$ and
$\tilde{l}$ in $S''_{1}$. The result  is the determinant,
which is the product of the eigenvalues
$\kappa$ that satisfy the equations
\bq
i \lambda_{1} \tilde{l} -p_{1}^{2} \tilde{q}&=& \kappa\, \tilde{q},\nonumber\\
i \lambda_{1} \tilde{q} -4\pi i \tilde{\lambda}_{2}(p_{0})\tilde{l}
&=& \kappa\, \tilde{l},
\eq
or
\be
\kappa^{2}+ \left(p_{1}^{2}+ 4\pi i \tilde{\lambda}_{2}(p_{0})\right)
\kappa +\lambda_{1}^{2}+ 4\pi i p_{1}^{2} \tilde{\lambda}_{2}(p_{0})
=0.
\ee
The eigenvalues depend on both $p_{0}\equiv p$, which is continuous, and on
$$
p_{1}=\frac{2\pi m}{p^{+}}
$$
with $m$ an integer. After the integration over $\tilde{q}$ and 
$\tilde{l}$, we end up with the factor
$$
(\det)^{-D/2}= \prod_{m}\prod_{p}\left(
 4\pi i \Big(\frac{2\pi m}{p^{+}}\Big)^{2}
\tilde{\lambda}_{2}(p)+\lambda_{1}^{2}\right)^{- D/2}
$$
The corresponding $Tr\ln$ is given by
\be
-\frac{ D}{2} Tr\ln= -\frac{D(\tau_{f}-\tau_{i})}{4\pi}\left(
\int d p\Big(\sum_{m} \ln\Big((\frac{2\pi m}{p^{+}})^{2}- \frac{i
\lambda_{1}^{2}}{4 \pi \tilde{\lambda}_{2}(p)}\Big)
+\frac{p^{+}}{\epsilon_{1}} \ln(\tilde{\lambda_{2}}(p))
\Big)\right).
\ee
In writing the right hand side of this equation, we have
factored $\tilde{\lambda}_{2}$ from the first term on the right,
and we incorporated it into the second term in the form of
$\ln(\tilde{\lambda}_{2})$. The coefficient of this term
involves a sum over m and appears to be divergent. However,
remembering that the $\sigma$ variable was discretized into
segments of length $\epsilon_{1}$, this sum is easily seen to be
equal to $p^{+}/\epsilon_{1}$.

We now collect various contributions to the effective action:
\be
S_{e}= S_{1}' + S_{F}- \frac{ D}{2} Tr\ln .
\ee
The terms that appear in this formula are given in eqs.(31),
(30) and (38). It is now easy to write down various classical
equations of motion that follow from this action. The equation
obtained by varying with respect to $\tilde{\lambda}_{2}$ is
\be
2\pi i D p^{+}(\tau_{f}-\tau_{i}) \tilde{\phi}_{2}(p)
-\frac{D}{2}\frac{\delta(Tr\ln)}{\delta(\tilde{\lambda}_{2}(p))}
=0,
\ee
or
\be
\tilde{\phi}_{2}(p)=\frac{\lambda_{1}^{2}}{(4\pi)^{3}
\tilde{\lambda}_{2}^{2}(p) s^{1/2}(p)}\coth\left(\frac{p^{+}}{2}
s^{1/2}(p)\right) -\frac{i}{8\pi^{2}\epsilon_{1}\tilde{\lambda}
_{2}(p)},
\ee
where, for convenience, we have defined
$$
s(p)= -\frac{i\lambda_{1}^{2}}{4\pi \tilde{\lambda}_{2}(p)}.
$$

Eq.(41) is the first fundamental equation of our
dynamical scheme; it establishes a relation between $\phi_{2}$
and $\lambda_{2}$. A second relation follows from varying the action
with respect to $\phi_{2}$:
$$
\frac{\delta(S_{e})}{\delta(\phi_{2})}=0.
$$
In this case, it turns out to be more convenient to stay in position
space and use $L$ instead of $\phi_{2}$ as the independent function.
 The connection between them is
$$
2 \phi_{2}(\tau)= L''(\tau),
$$
where we have assumed that $\tilde{\phi}_{2}(p)$ is an even function
of $p$. This assumption, although not crucial, simplifies matters.
It is consistent with all our equations and with
the approximate solutions given in Section 5.

Taking into account the contribution of $S_{1}'$ (eq.(31)) and
of $S_{F}$ (eq.(30)), we have
\be
D\, \lambda_{2}''(\tau)= \frac{2}{\epsilon_{1}} \frac{\delta
(\omega_{0})}{\delta(L(\tau))}.
\ee
We have now to compute the functional derivative 
 of $\omega_{0}$ with respect to $L$ from eq.(28).
Differentiating this equation gives
\be
\frac{\delta(\omega_{0})}{\delta(L(\tau))}= 2\pi ig^{2}\,
\frac{\delta(\tilde{K}(\omega_{0}))}{\delta(L(\tau))} \left(1 -
2\pi i g^{2} \tilde{K}'(\omega_{0})\right)^{-1},
\ee
and from eq.(25), we find
\be
\frac{\delta(\tilde{K}(\omega_{0}))}{\delta(L(\tau))}=
- \frac{D}{8\pi\epsilon}\Big(\frac{\pi}{\epsilon}\Big)^{D/2}
\exp\left(i \omega_{0} \tau +i D \phi_{1}\tau -\frac{D L(\tau)}{
4\epsilon}\right).
\ee
 
Eqs(42),(43) and (44) put together give us the second relation
between $\phi_{2}$, or equivalently $L$, and $\lambda_{2}$:
\be
\lambda''_{2}(\tau)=-\frac{i g_{0}^{2}
 \exp\Big(i\omega_{0} \tau
+ i D \phi_{1} \tau -\frac{ D L(\tau)}{4 \epsilon}\Big)}
{2 \epsilon \epsilon_{1}\left(1 + g_{0}^{2}
\int_{0}^{\infty} d\tau\,\tau \exp\Big(i \omega_{0} \tau+ i D
\phi_{1} \tau -\frac{D L(\tau)}{4 \epsilon}\Big)\right)},
\ee
where the dimensionless coupling constant $g_{0}$ is defined by
$$
g_{0}^{2}= g^{2}\left(\frac{\pi}{\epsilon}\right)^{D/2}.
$$

To the above equations, one has to add the
equations resulting from varying the action with respect to
$\lambda_{1}$ and $\phi_{1}$:
\bq
\lambda_{1} \phi_{1}&=& 4\pi \int d p \left(\tilde{\lambda}_{2}(p)
\tilde{\phi}_{2}(p)+\frac{i}{8 \pi^{2}\epsilon_{1}}\right),\nonumber\\
\lambda_{1}&=& 2 i \epsilon\int_{0}^{\infty}
 d\tau\,\tau\,\lambda_{2}''
(\tau).
\eq

 We now have a complete set of equations needed
to solve for the classical fields $\phi_{1,2}$ and $\lambda_{1,2}$.
We would like to stress that, apart from the large $D$ limit,
which is the basis of the mean field method, so far
 everything is exact.
Unfortunately, to make progress, we need to know $\omega_{0}$,
which is only given implicitly through eq.(28). Ultimately, it should
not be too difficult to find approximate solutions by numerical
methods. We will, however, leave this problem to future research.
Instead, in the next section, we will try
to extract as much information as we can from these equations
by analytic methods. Such information should be useful for
the eventual numerical work.

\vskip 9pt
\noindent{\bf 5. Consequences of the Dynamical Equations}
\vskip 9pt 

In the last section, we have derived a complete set of
equations for the dynamical variables $\lambda_{1,2}$
and $\phi_{1,2}$. This section will be devoted to the
investigation of the consequences of these equations.
First, for the convenience of the reader, we  collect the
the set of  equations to be investigated; namely, eqs.(28),(41) and
(45):
\bq
0 &=& \omega_{0}- i g_{0}^{2}
\int_{0}^{\infty} d\tau \exp\left(i \omega_{0} \tau+i D \phi_{1}\tau
- \frac{D}{4\epsilon} L(\tau)\right),\nonumber\\
\tilde{\phi}_{2}(p)&=&\frac{\lambda_{1}^{2}}{(4\pi)^{3}
\tilde{\lambda}_{2}^{2}(p) s^{1/2}(p)}\coth\left(\frac{p^{+}}{2}
s^{1/2}(p)\right) -\frac{i}{8\pi^{2}\epsilon_{1}\tilde{\lambda}
_{2}(p)},\nonumber\\
\lambda''_{2}(\tau)&=&-\frac{i g_{0}^{2}
 \exp\Big(i\omega_{0} \tau
+ i D \phi_{1} \tau -\frac{ D L(\tau)}{4 \epsilon}\Big)}
{2 \epsilon \epsilon_{1}\left(1 + g_{0}^{2}
\int_{0}^{\infty} d\tau\,\tau \exp\Big(i \omega_{0} \tau+ i D
\phi_{1} \tau -\frac{D L(\tau)}{4 \epsilon}\Big)\right)},
\eq
where
\be
L(\tau)=\int\frac{d p}{p^{2}}\left(2 -e^{i p\tau}- e^{- i p \tau}
\right) \tilde{\phi}_{2}(p),
\ee
which follows from eq.(26).

 We will now try to find an approximate
 solution to eqs.(47), and the approximation will be
based on a pole dominance model. This is a crude model, which is
at best valid only for large $\tau$, or equivalently, for small $p$.
We remind the reader that this region is of interest in probing
string formation (see the section following eq.(35)), so any
information gleaned is of value.
 Our starting point is the ansatz
\be
\tilde{\phi}_{2}\rightarrow \beta p^{2}
\ee
as $p\rightarrow 0$, where $\beta$ is a constant.
The basic idea is to cycle this limiting behavior through the dynamical
equations to check its consistency. We first need to restate it in the
position ($\tau$) space. Consider the $\tau\rightarrow \infty$ limit
of $L(\tau)$ from eq.(48). In the terms involving the exponentials,
this limit can be deduced from the well known Fourier relation:
Large $\tau$ is dominated by small $p$. Since the integrand is 
non-singular at $p=0$ by virtue of the ansatz on $\tilde{\phi}_{2}$,
 the exponential factors oscillate
to zero for $\tau\rightarrow\infty$, with the  result that
\be
L(\tau)\rightarrow L_{0},
\ee
 where $L_{0}$ is a constant. This constant is given by
\be
L_{0}=\int \frac{d p}{p^{2}}\, \tilde{\phi}_{2}(p).
\ee

If $L(\tau)$ goes to a constant as $\tau\rightarrow \infty$, the 
integral for $\tilde{K}(\omega)$ (eq.(25)) develops a pole in the variable
 $\omega$.
We will now make the assumption that this pole is a good approximation
for $\tilde{K}$ for small values of $\omega$ (pole dominance). We
therefore have
\be
\tilde{K}(\omega)\approx \frac{1}{2\pi}\Big(\frac{\pi}{\epsilon}\Big)
^{D/2}\exp\left(-\frac{D L_{0}}{4\epsilon}\right)
\frac{i}{\omega +D \phi_{1}}.
\ee
 Eq.(28) now becomes
\be
\omega_{0}^{2} + D\phi_{1} \omega_{0} +\bar{g}^{2}=0,
\ee
where we have defined
\be
\bar{g}^{2}= g_{0}^{2} \exp\left(-\frac{D L_{0}}{ 4 \epsilon}
\right).
\ee
The two solutions to this equation are
\be
\omega_{0}^{\pm}=-\frac{1}{2} D \phi_{1} \pm \frac{i}{2}\left(
  4 \bar{g}^{2}- D^{2} \phi_{1}^{2}  \right)^{1/2}.
\ee
We will investigate both solutions.
 
Next, we  evaluate $\lambda_{2}$ by replacing $L(\tau)$
by $L_{0}$ on the right hand side of eq.(45). As explained
above, this should be a good approximation in the large $\tau$,
small $p$ regime. Transforming into the momentum space, we
have
\bq
\tilde{\lambda}_{2}(p)&=& -\frac{\bar{g}^{2}}{4 \pi \epsilon
\epsilon_{1}}\frac{(\omega_{0}+D\phi_{1})^{2}}{\left((\omega_{0}
+D\phi_{1})^{2}- \bar{g}^{2}\right)(\omega_{0}+D\phi_{1} -p)}
\frac{1}{p^{2}}\nonumber\\
&\stackrel{p\rightarrow 0}{\rightarrow}&\pm \frac{i\bar{g}^{2}}
{4 \pi \epsilon \epsilon_{1} p^{2}}\left(
 4\bar{g}^{2}-(D\phi_{1})^{2} \right)^{-1/2}.
\eq
We note that the $1/p^{2}$ dependence of $\tilde{\lambda}_{2}$
for small $p$ is exactly what is needed for string formation
 (see eq.(35)). We can now substitute this result in eq.(41)
to determine the small p behavior of $\tilde{\phi}_{2}$, to
see whether it is consistent with our initial ansatz (49).
In the limit $p\rightarrow 0$, we indeed find that
\bq
\tilde{\phi}_{2}(p)&\rightarrow& \frac{i}{8\pi^{2}
\tilde{\lambda}_{2}(p)}\left(\frac{1}{p^{+}} -\frac{1}
{\epsilon_{1}}\right)\nonumber\\
&\rightarrow& {\rm const}\cdot  p^{2},
\eq
establishing consistency.

It is also important to establish the reality properties of various
variables within the context of our approximate solution. We
will initially assume that $\phi_{1}$ is pure imaginary and
$\bar{g}^{2}$ is real and positive. It then follows from eq.(56)
that, at least for small $p$, $\tilde{\lambda}_{2}$ is pure
imaginary. In the same small $p$ limit, eq.(57) tells us that
$\tilde{\phi}_{2}$ is real, and therefore, from eq.(51), so is
$L_{0}$. We are now able to verify one of our initial assumptions:
A real $L_{0}$ means that $\bar{g}^{2}$ is real and positive
(eq.(54)). It then follows from eq.(55) that $\omega_{0}^{\pm}$
are pure imaginary, and the corresponding energies $E_{0}^{\pm}$
(eq.(29)) are real, as they should be. Next, we will investigate
$\lambda_{1}$. It can be calculated from  the second eq.(46),
where $\lambda_{2}''$ is given by eq.(45), with $L(\tau)$
replaced by $L_{0}$. The result is
\be
\lambda_{1}=\frac{\bar{g}^{2}}{\epsilon_{1}}\frac{1}
{\bar{g}^{2}-\left(\omega_{0}+ D \phi_{1}\right)^{2}},
\ee
which shows that $\lambda_{1}$ is real. Finally, we go back
to the first equation (46): Since $\lambda_{1}$ and
$\tilde{\phi}_{2}$ are real, and $\tilde{\lambda}_{2}$ is
pure imaginary, it follows that $\phi_{1}$ is pure imaginary,
verifying the remaining initial assumption.

From the above analysis, we have seen that $\tilde{\lambda}_{2}(p)$
goes like $p^{2}$ for small $p$, which is the necessary condition
for string formation. However, in addition, the square of the
slope parameter
$\alpha$ (eq.(35)) must be real and positive. Using eq.(56), this
parameter is given by
\be
\alpha^{2}=\frac{\lambda_{1}^{2}}{16 \pi i p^{2}\tilde{\lambda}
_{2}(p)}=\mp\frac{\epsilon\epsilon_{1} \lambda_{1}^{2}}
{4 \bar{g}^{2}}\left(4 \bar{g}^{2}- D^{2}\phi_{1}^{2}\right)^{1/2},
\ee
and remembering that $\lambda_{1}$ is real and $\phi_{1}$ is
pure imaginary, $\alpha^{2}$ is positive for the lower sign.
This means that in eq.(55) for $\omega_{0}$, we must also choose
the lower sign. This is then the string forming solution,
 and from the same equation,
 we see that the corresponding energy $E_{0}^{-}$ is positive.
However, the other solution, corresponding to the upper sign,
has lower energy, since $E_{0}^{+}$ is negative, so that it
will actually dominate the worldsheet path integral, and
not describe a string. The state that looks like string
is at best meta-stable and more likely completely unstable.
This is, of course,
not unexpected, since $\phi^{3}$ is an inherently unstable
theory. We expect a more realistic theory, such as a 
non-Abelian gauge theory, to retain the string forming solution
but to be free of the instability. As explained in the 
introduction, our goal in studying $\phi^{3}$ is to develop
the tools needed to attack more realistic but also more
complicated theories in a simpler setting. We find it
encouraging that, despite the instability, we find a signal
for string formation.

The results discussed above were based on the pole dominance
approximation. As we have argued earlier, this is probably a 
reasonable approximation for small $p$. In fact, the $1/p^{2}$
dependence of $\tilde{\lambda}_{2}$ for small $p$, which
leads to string formation, appears to be generic. This is
easily seen from eq.(45): In momentum space, the double
derivative on $\lambda_{2}$ turns into a factor of $p^{2}$,
and unless the right hand side of this equation vanishes
accidentally at $p=0$, $\tilde{\lambda}_{2}$ must have
a $1/p^{2}$ dependence. On the other hand, we cannot expect
the pole dominance model to be valid for large $p$. In
fact, we show below that the large $p$ regime is quite
different from the small $p$ regime.

We will try to determine the large $p$ behavior of the
fields through self consistency.  We start with the ansatz
$$
\tilde{\phi}_{2}(p)\rightarrow b |p|,
$$
for $p\rightarrow\infty$, where $b$ is a constant, or,
equivalently
$$
\phi_{2}(\tau)\rightarrow -\frac{2 b}{\tau^{2}}
$$
as $\tau \rightarrow 0$. It then follows from
$$
2\phi_{2}(\tau)= L''(\tau)
$$
that
$$
L(\tau)\rightarrow 4 b \ln(\tau),
$$
in the same limit. Eq.(45) then gives
$$
\lambda_{2}(\tau)\rightarrow {\rm const}\cdot 
\tau^{(2- {b D}/{\epsilon})}
$$
again, as $\tau\rightarrow 0$. The motivation for the original
ansatz was in fact to obtain a power behavior for $\lambda_{2}$,
which translates into
$$
\tilde{\lambda}_{2}(p)\rightarrow {\rm const}\cdot 
|p|^{{b D}/{\epsilon} - 3}
$$
for large $p$ in momentum space. Now, from eq.(41), we find that
$$
\tilde{\phi}_{2}(p)\rightarrow {\rm const}\cdot \tilde{\lambda}(p)^{-3/2}
\rightarrow {\rm const}\cdot |p|^{{9}/{2}-{3 b D}/{2 \epsilon}},
$$
and comparing this with the initial ansatz, we see that it is
consistent if
$$
\frac{b D}{\epsilon}=\frac{7}{3}.
$$
 Admittedly,
this is a crude analysis, but if we accept it, we have asymptotic limit
\be
\tilde{\lambda}_{2}(p)\rightarrow {\rm const}\cdot |p|^{-2/3},\;\;\;
\tilde{\phi}_{2}(p)\rightarrow {\rm const}\cdot |p|,
\ee
for  $p\rightarrow \infty$. This is quite different from the
small $p$ behavior
\be
\tilde{\lambda}_{2}(p)\rightarrow {\rm const}\cdot 1/p^{2},\;\;\;
\tilde{\phi}_{2}(p)\rightarrow {\rm const}\cdot p^{2}.
\ee

We end this section with a short discussion of the need for
a third cutoff. The asymptotic behavior derived above
shows that the integral for $L$ (eq.(48), as well as the
integral for $\lambda_{1} \phi_{1}$ (eq.(46)), are divergent
at large $p$. The integrals have to be regulated, say, by a 
factor of the form
$$
\exp(-\epsilon_{2} p^{2}),
$$
where $\epsilon_{2}$ is a third cutoff, in addition to $\epsilon$
and $\epsilon_{1}$. This new cutoff is probably not independent
of $\epsilon$. The argument goes as follows: 
Instead of an exponential cutoff, let us latticize the $\tau$
direction, with a lattice spacing $\Delta \tau$. From (6), we see
that one lattice step $\Delta\tau$ corresponds to a factor of
$$
\exp\left(-\Delta\tau\, {\bf p}^{2}/ 2 p^{+}\right).
$$
Comparing this with the cutoff $\exp(-\epsilon k^{2})$ (eq.(15)),
we can identify
\be
\Delta\tau \approx \epsilon p^{+}.
\ee
On the other hand, there is a  rough relation between $\Delta\tau$
and $\epsilon_{2}$:
$$
\epsilon_{2}\approx (\Delta\tau)^{2},
$$
and therefore, it follows that
\be
\epsilon_{2}\approx (\epsilon p^{+})^{2}.
\ee

In view of this connection between $\epsilon$ and $\epsilon_{2}$,
it is tempting to conjecture that one of them is redundant. In fact,
in eqs.(46) and (47), $\epsilon$ and $\epsilon_{1}$ do not regulate any
divergent integrals; they merely appear as parameters. One can
easily eliminate the singular dependence on these parameters by
the following scaling of the fields:
\bq
\phi_{2}&\rightarrow& \epsilon \phi_{2},\;\;\;
\lambda_{2} \rightarrow \frac{\lambda_{2}}{\epsilon \epsilon_{1}},
\nonumber\\
 \lambda_{1}&\rightarrow& \frac{\lambda_{1}}{\epsilon_{1}},
\;\;\;\phi_{1}\rightarrow \phi_{1},\;\;\;\omega_{0}\rightarrow
\omega_{0}.
\eq
After this scaling, as far as the equations are concerned, one can let
$$
\epsilon\rightarrow0,\;\;\;\epsilon_{1}\rightarrow 0,
$$
while keeping the dimensionless coupling constant $g_{0}$ fixed.
Let us see how some physical quantities behave
under this scaling. As indicated above, $\omega_{0}$ and therefore
$E_{0}$ are independent of $\epsilon$ and $\epsilon_{1}$, whereas
the slope parameter $\alpha^{2}$ scales as $\epsilon/\epsilon_{1}$
(eq.(35)).
Keeping this ratio finite as each cutoff parameter goes to zero
would avoid any singular behavior. So it appears that at least
in the leading order of the mean field approximation, both
$\epsilon$ and $\epsilon_{1}$ are redundant.
 In contrast, as pointed out
earlier, the cutoff $\epsilon_{2}$ cannot be dispensed with;
it is really needed to regulate integrals over $p$ in equations
(46) and (48).
 Clearly, the problems involving cutoff dependence and
 renormalization must await a better understanding
of the solution to eqs.(47).

\vskip 9pt
\noindent{\bf 6. Conclusions}
\vskip 9pt

The main result of this article is the derivation of a set of
equations for the sum of all planar graphs the $\phi^{3}$
field theory, subject to the mean field or large $D$
approximation. Apart from this single approximation, the
treatment is exact. This is in contrast to our earlier paper
\cite{bardakcitmean}, where the 
same problem was considered, but also further and
somewhat questionable approximations were made. The present
treatment is therefore a definite improvement over the one 
presented in \cite{bardakcitmean}.

The equations derived in this article are somewhat involved
but they appear treatable by numerical methods. As a preliminary
step, we have studied them in the large and small momentum
regimes. The latter is especially important since it is
relevant to string formation. We find that a string forming
solution is consistent with our equations, although this
solution is unstable. Given that we are dealing with an
unstable theory to start with, we still find this result
encouraging.

Much still remains to be done. A thorough numerical analysis
of the equations derived here should be carried out. Since
we do not necessarily expect to find interesting results
for the $\phi^{3}$ model, the methods developed in this
article should be applied to more physical theories, whose world 
sheet formulations are already available \cite{thornsheet,gudmundssontt}. 
It is also important
to go to next to leading order in the large $D$ expansion, both as
a check on the leading order, and also to investigate questions
such as Lorentz invariance.
\vskip 9pt
\noindent{\bf Acknowledgments}
\vskip 9pt
We would like to thank Jeff Greensite for valuable discussions.
This work was supported in part
 by the Director, Office of Science,
 Office of High Energy and Nuclear Physics, 
 of the U.S. Department of Energy under Contract 
DE-AC03-76SF00098, in part by the National Science Foundation Grant
22386-13067-44-X-PHHXM, and in part by the Department of Energy
under Grant No. DE-FG01-97ER-41029

\newpage
\noindent{\bf Appendix. Contribution of the Ghost Sector}
\vskip 9pt

In this appendix, we will discuss the changes resulting from
taking into account the ghost fields $b$ and $c$. First of all,
an additional factor has to be added to $K$ given by (14):
$$
K\rightarrow K \times K_{g},
$$
where,
\be
K_{g}(\sigma,\tau_{2},\tau)= \int D \bar{b} \int D \bar{c}
\exp\left(i \int_{\tau_{2}}^{\tau} d \tau' \Big(\bar{b}(\sigma,\tau')
\cdot b(\sigma,\tau')+ \bar{c}(\sigma,\tau')\cdot c(\sigma,\tau')
\Big)\right).
\ee
Here, $\bar{b}$ and $\bar{c}$ are lagrange multipliers that 
enforce the boundary conditions (5) on solid lines. Moreover,
there is a ghost contribution to the action(eq.(28)):
$$
S\rightarrow S+ S_{g},
$$
where,
\be
S_{g}=\int_{0}^{p^{+}}\int d\tau\left(b'\cdot c' + \kappa_{1}
(\bar{b}\cdot b - D \eta_{1})+ \kappa_{2}(\bar{c}\cdot c - D \eta_{2})
\right).
\ee

In this expression, the first term comes from eq.(2), and in
the rest of the terms, the lagrange multipliers $\kappa_{1,2}$
set the expectation value of $\bar{b}\cdot b$ and $\bar{c}\cdot c$
equal to $\eta_{1,2}$. As before, this is in preparation for taking
the large $D$ limit: Vector valued fields $\bar{b}$ and $\bar{c}$
are traded for scalar valued fields $\kappa_{1,2}$ and $\eta_{1,2}$.

Next, as in eq.(22), we use translation invariance to replace the
fields $\kappa_{1,2}$ and $\eta_{1,2}$ by constants independent
of $\sigma$ and $\tau$. As a result, we can set
\be
K_{g}\rightarrow \exp\left(i D (\eta_{1}+\eta_{2})(\tau -\tau_{2})
\right),
\ee
and eq.(18) gets replaced by
\be
K\rightarrow (\pi/\epsilon)^{D/2} \exp\left(i D (\tau_{2} -\tau_{1})
(\phi_{1} +\eta_{1} + \eta_{2}) -\frac{D}{4 \epsilon} L(\tau_{2}
-\tau_{1}) \right).
\ee
We see that as far as $K$ and therefore $F$ is concerned, all that 
the ghosts have done is to shift $\phi_{1}$ by $\eta_{1}+\eta_{2}$.
As a result, eqs.(25) through (28) still continue to be valid, with
the proviso that $\phi_{1}$ be replaced by the shifted $\phi_{1}$.
These shifts do not change anything done in this paper, since we
 have never made any use of the actual value of $\phi_{1}$,
and that is the reason why we have suppressed the ghost sector
in the main body of the paper.
 We expect them to become important in the next non-leading
order in $D$.

Although we do not need it, for the sake of completeness, we will
carry out the functional integrations over $b$, $c$, $\bar{b}$
and $\bar{c}$ in $S_{g}$ (eq.(62)). The resulting $Tr\ln$ is very
simple:
\be
Tr \ln_{g}= p^{+} (\tau_{f}-\tau_{i}) \frac{C}{16}
\kappa_{1}^{2} \kappa_{2}^{2},
\ee
where $C$ is a cutoff dependent constant. The reason for this
simple answer is that the term $b'\cdot c'$ in (62), the only term
that ``propagates'' the ghosts, does not contribute to the determinant.
This is an example of the non-dynamical, or equivalently, non-propagating
ghosts discussed in \cite{bardakcitmean}. In the absence of propagation, the determinant,
or the $Tr\ln$ has purely algebraic dependence on the $\kappa$'s.
In any case, we see that the ghost contribution to the action is
relatively trivial.


\end{document}